\documentclass[aps,prd,a4paper,preprintnumbers,tightenlines,superscriptaddress,
twocolumn,showpacs,final,nofootinbib,floatfix]{revtex4-1}
\pdfoutput=1
\usepackage{graphicx}
\usepackage{longtable}
\usepackage{amsmath}
\usepackage{amssymb}
\usepackage{hyperref}
\usepackage{dcolumn}
\usepackage{bm}
\usepackage{multirow}
\usepackage{cleveref}
\usepackage{bm}
\usepackage{color}

\newcommand{\be}{\begin{eqnarray}}
\newcommand{\ee}{\end{eqnarray}}
\newcommand{\mpl}{M_{\rm {pl}}}

\newcommand{\dd}{\, {\rm d}}
\newcommand{\gsim}{\;\mbox{\raisebox{-0.5ex}{$\stackrel{>}{\scriptstyle{\sim}}$}
}\;}
\newcommand{\lsim}{\;\mbox{\raisebox{-0.5ex}{$\stackrel{<}{\scriptstyle{\sim}}$}
}\;}
\def\eea{\end{eqnarray}}
\def\bea{\begin{eqnarray}}

\newcommand{\ve}{\varepsilon}

\newcommand{\BI}{\bar{I}}

\newcommand{\ccc}{_{\rm c}}

\newcommand{\U}{\Upsilon}

\newcommand{\com}{\mathcal{C}}

\newcommand{\nm}{{\mu\nu}}

\newcommand{\rpp}{r'}
\newcommand{\GN}{G_{\rm N}}

\newcommand{\dn}{\delta\nu}
\newcommand{\dl}{\delta\lambda}
\newcommand{\oo}{\mathcal{O}}

\newcommand{\lag}{\mathcal{L}}
\newcommand{\obar}{\omega}
\newcommand{\uo}{\Upsilon_1}
\newcommand{\z}{\zeta}

\newcommand{\ut}{\Upsilon_2}
\newcommand{\pdot}{v_0}

\allowdisplaybreaks
\begin{document}
\title{Towards Strong Field Tests of Beyond Horndeski Gravity Theories}
\author{Jeremy Sakstein}
\email[Email: ]{sakstein@physics.upenn.edu}
\affiliation{Center for Particle Cosmology, Department of Physics and Astronomy, University of Pennsylvania, 209 S. 33rd St., Philadelphia, PA 19104, USA}
\affiliation{Institute of Cosmology and Gravitation, University of Portsmouth, Portsmouth, PO1 3FX, UK}
\author{Eugeny Babichev}
\email[Email: ]{eugeny.babichev@th.u-psud.fr}
\affiliation{Laboratoire de Physique Th\'eorique, CNRS, Univ.~Paris-Sud, Universit\'e Paris-Saclay, 91405 Orsay, France}
\author{Kazuya Koyama}
\email[Email: ]{kazuya.koyama@port.ac.uk}
\affiliation{Institute of Cosmology and Gravitation, University of Portsmouth, Portsmouth PO1 3FX, UK}
\author{David Langlois}
\email[Email: ]{langlois@apc.univ-paris7.fr}
\affiliation{Laboratoire APC -- Astroparticules et Cosmologie, Universit\'e Paris Diderot Paris 7, 75013 Paris, France}
\author{Ryo Saito}
\email[Email: ]{rsaito@yukawa.kyoto-u.ac.jp}
\affiliation{Center for Gravitational Physics, Yukawa Institute for Theoretical Physics, Kyoto University, 606-8502, Kyoto, Japan}

\begin{abstract}
Theories of gravity in the beyond Horndeski class encompass a wide range of scalar-tensor theories that will be tested on cosmological scales over the coming decade. In this work, we investigate the possibility of testing them in the strong-field regime by looking at the properties of compact objects---neutron, hyperon, and quark stars---embedded in an asymptotically de Sitter space-time, for a specific subclass of theories. We extend previous works to include slow rotation and find a relation between the dimensionless moment of intertia, ($\BI=Ic^2/\GN M^3$), and the compactness, $\com=\GN M/Rc^2$ (an $\BI$--$\com$ relation), independent of the equation of state, that is reminiscent of but distinct from the general relativity prediction. Several of our equations of state contain hyperons and free quarks, allowing us to revisit the hyperon puzzle. We find that the maximum mass of hyperon stars can be larger than $2M_\odot$ for small values of the beyond Horndeski parameter, thus providing a resolution of  the hyperon puzzle based on modified gravity. Moreover, stable quark stars exist when hyperonic stars are unstable, which means that the phase transition from hyperon to quark stars is predicted just as in general relativity, albeit with larger quark star masses. Two important and potentially observable consequences of some of the theories we consider are the existence of neutron stars in a range of masses significantly higher than in GR, and $\BI$--$\com$ relations that differ from their GR counterparts. In the former case, we find objects that, if observed, could not be accounted for in GR because they violate the usual GR causality condition. We end by discussing several difficult technical issues that remain to be addressed in order to reach more realistic predictions that may be tested using gravitational wave searches or neutron star observations.
\end{abstract}

\maketitle

\section{Introduction}

Since its inception over a century ago, general relativity (GR) has proved a phenomenal success, passing tests with ever-increasing precision in a variety of different environments \cite{Baker:2014zba} from the solar system \cite{Will:2004nx} to the laboratory \cite{Burrage:2016bwy}, and, recently, gravitational waves consistent with its predictions have been observed \cite{Abbott:2016blz,Abbott:2016nmj}. Despite this, there are relatively few tests of gravity in the strong field regime\footnote{Although, see \cite{Damour:1998jk} and references therein for some tests of conformal scalar-tensor theories.}. Indeed, whilst the recent LIGO observations are consistent with GR's predictions, the experiment is not yet able to distinguish between GR and non-GR polarisations i.e. scalar and vector modes  \cite{TheLIGOScientific:2016src} and, similarly, constraints on alternative theories of gravity are not possible due to some limitations of theoretical predictions. 
In particular, since the detected events correspond to a black hole binary coalescing, one major obstacle is the powerful no-hair theorem \cite{Bekenstein:1971hc,Bekenstein:1995un} (see \cite{Jacobson:1999vr,Weinberg:2001gc,Herdeiro:2014goa,Babichev:2016rlq} and references therein for some notable exceptions), stating that black holes are fully described by their mass, charge, and angular momentum, and therefore behave in an identical manner to GR\footnote{This is somewhat of a generalisation, it may be that the solutions for the metric are identical but different dynamics lead to different gravitational wave signals. The lack of numerical simulations is the limiting factor in this case, although some progress has been made, e.g. \cite{Berti:2013gfa}.}. 

Neutron stars (and other more speculative compact objects such as hyperon stars, quark stars, and boson stars) are another promising avenue for testing gravity in the strong field regime since the no-hair theorems apply strictly to vacuum solutions. Neutron star solutions have been obtained for several different alternative gravity theories (see \cite{Berti:2015itd} and references therein).
Unfortunately, tests of gravity using these objects are highly degenerate with baryonic physics due to large theoretical uncertainties in the nuclear equation of state (EOS) \cite{Lattimer:2012nd}. 
 Universal (or approximately universal) relations between different compact star properties (such as the I-Love-Q relation) have been found \cite{Yagi:2016bkt}, and, just recently, \cite{Breu:2016ufb} have identified a similar relation between the dimensionless moment of inertia ($\bar{I}=Ic^2/G^2M^3$) and the compactness $\com=GM/Rc^2$. Such relations are useful for testing gravity precisely because any observed deviation cannot be attributed to the equation of state, and, furthermore, alternative theories that predict different relations can be constrained by their measurement. 

One of the main drivers of modified gravity research is the cosmological constant problem and the elusive nature of dark energy \cite{Copeland:2006wr,Clifton:2011jh,Joyce:2014kja,Koyama:2015vza,Bull:2015stt}. The modern approach to scalar-tensor theories relies on the framework of \emph{beyond Horndeski} theories \cite{Gleyzes:2014dya,Gleyzes:2014qga} (see also \cite{Zumalacarregui:2013pma,Deffayet:2015qwa,Crisostomi:2016tcp}). This refers to a very broad class of models that are free from the Ostrogradski ghost instability, and are therefore seen as healthy cosmological theories, although they may not be effective field theories in the quantum field theory sense. Note that beyond Horndeski theories are included in a more general class of scalar-tensor theories that have higher order equations of motion and contain at most three propagating degrees of freedom \cite{Langlois:2015cwa,Langlois:2015skt,Crisostomi:2016czh,deRham:2016wji,Achour:2016rkg,Motohashi:2016ftl,Ezquiaga:2016nqo,BenAchour:2016fzp}.

Beyond Horndeski theories are cosmological competitors to the $\Lambda$CDM model (GR and a cosmological constant); they admit late-time de Sitter attractors without the need for a cosmological constant\footnote{Like most alternative gravity theories, one must still explain why the contribution from the cosmological constant is zero from the outset.}. Furthermore they make novel predictions for astrophysical objects such as stars and galaxies \cite{Kobayashi:2014ida,Koyama:2015oma,Saito:2015fza,Sakstein:2015zoa,Sakstein:2015aac,Jain:2015edg,Sakstein:2016ggl,Sakstein:2016lyj}, and, somewhat amazingly, the parameters that control said effects are precisely those that control the linear cosmology of the theories that will be probed by upcoming missions. Small-scale tests of the theory therefore have the power to constrain cosmological modifications of gravity, which has motivated a recent effort focused on understanding the strong-field regime of this theory. 

In a previous paper \cite{Babichev:2016jom}, we have made some progress towards this using a simple model where the deviations from GR are characterised by a single dimensionless parameter $\U$. We performed a preliminary investigation into the properties of neutron stars by deriving the modified Tolman-Oppenheimer-Volkoff (mTOV) system of equations governing the static equilibrium structure and solved them for two realistic equations of state to find the mass-radius relation. The purpose of this work is to build on this in an effort to identify potential observables, and, more importantly, to demonstrate that compact objects can show significant deviations from the GR predictions. To this end, we extend our previous formalism to the case of slow rotation and integrate the equations for 32 equations of state. These equations of state have all been computed using different calculational methods such as Hartree-Fock, relativistic mean field theory, Skyrme models, etc. and cover the full spectrum of formalisms used in the literature; we describe them in detail in Appendix \ref{app:EOS}. We use these equations of state to compute the $\bar{I}$--$\com$ relation, and make general statements about neutron star properties. We also investigate compact stars containing hyperons and quarks. Our findings can be summarised as follows:
\begin{itemize}
\item We generically find neutron stars with masses larger than $2M_\odot$ when $\U\lsim-0.03$ and masses of order $3M_\odot$ or larger are typically found for the stiffer equations of state favoured currently. The largest mass neutron star thus far observed has a mass $2.01\pm0.04M_\odot$ \cite{Antoniadis:2013pzd}.
\item We find a universal $\bar{I}$--$\com$ relation which has a similar shape to that found in GR. The differences between GR and beyond Horndeski theory is larger than the scatter due to different equations of state when $\U\lsim -0.03$, indicating that a measurement of this relation could discriminate between different theories. Such a measurement may be possible within the next decade.
\item Hyperon stars with masses $\gsim2M_\odot$ can be obtained for beyond Horndeski parameters not yet constrained by observations, thereby providing a new solution to the hyperon puzzle.  
\item Massive hyperon stars can transition into quark stars with masses in excess of $2M_\odot$.
\end{itemize}

One important difference between GR and beyond Horndeski theories is that the asymptotics are important in determining the small-scale physics. Indeed, if the cosmological time-derivative of the scalar is zero then the resultant compact objects are identical to those predicted by GR. 
In order to account for this in a fully relativistic manner it has been necessary to pick a simple subset of beyond Horndeski theories that admit exact de Sitter solutions and consider compact objects embedded in such space-times\footnote{The simplest model includes a cosmological constant \cite{Babichev:2016jom} but in this work we present a more general class of models that does not require such a term and is therefore more natural from a theoretical point of view.}. We elucidate the technical difficulties involved in going beyond these simple models and discuss various approaches one could take to address these in future studies; such studies lie outside the scope of the present work. Having demonstrated here that large deviations from GR can be found in these theories, a study extending the results presented here is clearly warranted before experiments such as the LIGO/Virgo collaboration can begin to search for signatures of beyond Horndeski theories \cite{Yunes:2016jcc}. We end this paper by discussing how one might go about performing such calculations.

This paper is organised as follows: In section \ref{sec:modcos} we briefly review the pertinent aspects of beyond Horndeski theories and introduce the specific model we will study. The derivation of the mTOV equations as well as the equation governing slow rotation is long and technical, and, for this reason, we present their derivation in Appendix \ref{sec:deriv}. In section \ref{sec:NS} we study neutron and more exotic stars. We discuss the maximum neutron star mass, present the modified $\bar{I}$--$\com$ relations, and examine the hyperon puzzle. In section \ref{sec:disc} we discuss our findings in the context of testing GR; we also discuss the technical challenges associated with constructing more realistic models, placing emphasis on the hurdles future studies would need to focus on overcoming. We conclude in section \ref{sec:concs}.

\section{Beyond Horndeski Theories}\label{sec:modcos}

Beyond Horndeski theories \cite{Gleyzes:2014dya,Gleyzes:2014qga} are a very broad class of scalar-tensor theories that exhibit interesting properties that make them perfect paragons for alternative gravity theories. The effects of modified gravity are hidden, or \emph{screened}, in the solar system by the Vainshtein mechanism \cite{Babichev:2013usa} and so classical tests of gravity based on the parameterised post-Newtonian (PPN) framework are automatically satisfied\footnote{This is the case for most Vainshtein screened theories although it has yet to be shown in full generality. In particular, it has not been investigted for theories that exhibit Vainshtein breaking with the exception of the Eddington light bending parameter $\gamma$ (in all theories \cite{Kobayashi:2014ida,Koyama:2015oma}) and $\beta$ (only simple theories such as the one presented in \cite{Babichev:2016jom} and extended here or those which admit exact Schwarzchild-de Sitter solutions such as theories in the \emph{three graces} class \cite{Babichev:2016kdt}). See the discussion in section \ref{sec:concs}.} but novel deviations from GR, often referred to as \emph{Vainshtein breaking}, are seen inside astrophysical bodies so that the equations of motion for the weak-field metric potentials defined by
\begin{equation}\label{eq:iso}
\dd s^2 =-(1+2\Phi)\dd t^2 + 
(1-2\Psi)
\, \delta_{ij}\dd x^i\dd x^j 
\end{equation}
are modified to \cite{Kobayashi:2014ida,Koyama:2015oma,Saito:2015fza,Sakstein:2016ggl}
\begin{align}
\label{Upsilons}
\frac{\dd\Phi}{\dd r}&=\frac{\GN M(r)}{r^2}+\frac{\uo \GN}{4}\frac{\dd^2M(r)}{\dd r^2}\\
\frac{\dd\Psi}{\dd r}&=\frac{\GN M(r)}{r^2}-\frac{5\ut\GN}{4r}\frac{\dd M(r)}{\dd r}.
\end{align}
The dimensionless parameters $\Upsilon_i$ characterise deviations from GR 
of the beyond Horndeski type. They are directly related to the parameters appearing in the effective description of dark energy that controls the linear cosmology of beyond Horndeski theories \cite{Gleyzes:2013ooa,Bellini:2014fua,Gleyzes:2014rba} via \cite{Saito:2015fza,Sakstein:2016ggl}:
\begin{align}\label{eq:EFT}
\Upsilon_1&=\frac{4\alpha_H^2}{c_T^2(1+\alpha_B)-\alpha_H-1}\quad\textrm{and}\nonumber\\ \Upsilon_2 &= \frac{4\alpha_H(\alpha_H-\alpha_B)}{5(c_T^2(1+\alpha_B)-\alpha_H-1)}.
\end{align}
The coefficients $\alpha_i$ (see \cite{Gleyzes:2014rba} for their definitions) will be constrained by future cosmological surveys aimed at testing the structure of gravity on large scales \cite{Alonso:2016suf} and so any constraints from small scale probes are complimentary to these and may provide orthogonal bounds. 
When $\alpha_H=0$, as is the case in GR but also in Horndeski theories, the parameters $\Upsilon_i$ vanish. 
Currently, $\uo$ is constrained to lie in the range $-0.48<\uo<0.027$ \cite{Sakstein:2015zoa,Sakstein:2015aac,Jain:2015edg,Sakstein:2016lyj} using stellar tests whilst $\ut$ is only weakly constrained by galaxy cluster tests \cite{Sakstein:2016ggl}. This is partly due to the need for relativistic systems that probe $\Psi$ and so the strong field regime is perfect for placing more stringent constraints. The upper bound on $\uo$ is free of the technical ambiguities related to the strong field regime (non-relativistic stars are used) and so, for this reason, we focus exclusively on the case $\uo<0$ in this work.

\subsection{Model}

The specific action we will consider is
\begin{equation}
\label{eq:action}
S=\int d^4x\,  \sqrt{-g}\, \left[
\mpl^2\left(\frac{R}{2}-k_0\Lambda\right)-\mathcal{L}_2+f_4\lag_{4,{\rm bH}}\right]\,,
\end{equation}
with
\begin{align}
\mathcal{L}_2&= k_2X +\frac{\z}{2} X^2 ,\\
\mathcal{L}_{4,{\rm bH}}&=-X\left[(\Box\phi)^2-(\phi_{\nm})^2\right]+2\phi^\mu\phi^\nu\left[\phi_\nm\Box\phi-\phi_{\mu\sigma}\phi^{\sigma}_{\,\,\nu}\right],
\end{align}
with X=$\phi_{\mu}\phi^{\mu}$; $\Lambda$ is a (positive) cosmological constant and $k_2$, $\zeta$, $f_4$, and $k_0$ are constant coefficients. 
The above action belongs to the family of the beyond Horndeski theory (``three Graces''), 
which admits exact Schwarzschild-de-Sitter outside a star, with a time-dependent scalar field \cite{Babichev:2016kdt}\footnote{Exact Schwarzschild-de Sitter solutions with a time-dependent scalar field also exist in the Horndeski theory~\cite{Babichev:2013cya,Kobayashi:2014eva}.}.
Our previous model \cite{Babichev:2016jom} had $k_0=1$ and $\zeta=0$ but in order to have a model that does not contain a bare cosmological constant we will instead set $k_0=0$. This action is by no means the most general beyond Horndeski action but contains simple models that exhibit Vainshtein breaking. 
The $\mathcal{L}_{4,{\rm bH}}$ term corresponds to a covariantisation of the quartic galileon.

The derivation of the mTOV equations, which follows the procedure set out in \cite{Babichev:2016jom}, is rather long and technical, as is the derivation of the new differential equation we will solve in this work. For this reason, the main details can be found in Appendix \ref{sec:deriv} but here we will outline the important points briefly. We will follow the method of Hartle and Thorne \cite{Hartle:1967he,Hartle:1968si} and write the metric as 
\begin{align}\label{eq:HT}
\dd s^2 &= -e^{\nu(r)}\dd t^2 + e^{\lambda(r)}\dd r^2 + r^2\left(\dd \theta^2 +\sin^2\theta\dd\phi^2\right)\nonumber\\& - 2\ve(\Omega-\omega(r))r^2 \sin^2\theta \dd t \dd\phi.
\end{align}
This describes the geometry of a space-time containing a slowly rotating  star with angular velocity $\Omega$. The slow rotation is enforced by using a small dimensionless book-keeping parameter $\ve\ll1$. 

At zeroth order, i.e. for $\ve=0$,  the metric is static and spherically symmetric. It is convenient to decompose each metric potential, as well as the scalar field, into a cosmological contribution, corresponding to the de Sitter solution (in Schwarzschild coordinates), and a contribution sourced by the star~\cite{Babichev:2016jom}: 
\begin{align}
\label{asymp_dS}
 \nu&= \ln\left(1-H^2r^2\right) + \delta\nu(r)\\
  \lambda&= - \ln\left(1-H^2r^2\right)+\delta\lambda,\\
 \phi &= v_0t+\frac{v_0}{2H}\ln\left(1-H^2r^2\right) + \varphi(r).\label{eq:phids}
\end{align}
The constant $\pdot=\dot{\phi}$ is the time-derivative of the cosmological scalar field that satisfies the Friedmann equations. Substituting all this into the equations of motion with the star described by a perfect fluid, and eliminating the scalar $\varphi$ in the sub-Horizon limit (see \cite{Babichev:2016jom} for the details), one obtains a system of differential equations for $\dn$ and $\dl$ (the mTOV system) that can be solved given an equation of state. These equations, which  are given in Appendix \ref{sec:deriv}, are identical to those derived in \cite{Babichev:2016jom}. 
By comparing these equations with  (\ref{Upsilons}),   one finds that
\begin{equation}
\uo=\ut\equiv\U=\frac{s-2}{3},
\end{equation}
where $\z=6\mpl^2H^2(1-s)/\pdot^4$ (see Appendix \ref{sec:deriv}). 

At order $\oo(\ve)$, we obtain an additional equation for the extra metric function $\omega$,  of the form
\begin{equation}\label{eq:roteqn}
\omega''= K_1(P,\rho,\dl,\dn,\U)\omega'+K_0(P,\rho,\dl,\dn,\U)\omega,
\end{equation}
where $K_0$ and $K_1$ are complicated functions given in Appendix \ref{sec:deriv}. Once the mTOV equations for $\dn$ and $\dl$ have been solved, one can use them as inputs to solve this equation for $\omega$. 
As shown in Appendix \ref{sec:deriv}, outside the star the equation for $\omega$ reduces to the one predicted by GR so that
\begin{equation}\label{eq:J}
\obar=\Omega-2\frac{\mathcal{J}}{r^3}
\end{equation}
outside the star \cite{Hartle:1967he}. Here, $\mathcal{J}$ is the angular momentum from which one can extract the moment of inertia $I=\mathcal{J}/\Omega$. Note that at large distances one has $\omega\rightarrow\Omega$ 
so that the space-time is asymptotically de Sitter. Equation \eqref{eq:J} implies that
\begin{equation}\label{eq:Iexpr}
I=\frac{R^4}{2}\frac{\omega'(R)/\omega(R)}{3+R\omega'(R)/\omega(R)},
\end{equation}
where $R$ is the radius of the star. Equation \eqref{eq:roteqn} is homogeneous and so once a solution is found, one is free to rescale $\omega$ by a constant factor to find another solution. The expression \eqref{eq:Iexpr} for $I$  is invariant under such a rescaling and so we can set the central value $\omega(0)=\omega\ccc=1$ without loss of generality. Furthermore, spherical symmetry imposes that $\omega'(0)=0$. 
We solve Equation \eqref{eq:roteqn} with these boundary conditions at the centre and find $\omega(R)$ and $\omega'(R)$ at the surface of the star. We then compute $I$ using Equation \eqref{eq:Iexpr}. 

\section{Compact Objects}\label{sec:NS}

In our previous paper \cite{Babichev:2016jom}, we used two equations of state to investigate the properties of neutron stars in beyond Horndeski theories. In this work, we use 32 equations of state that have been proposed in the literature; similar investigations have been performed for other modified gravity theories using some of these equations of state \cite{Cisterna:2015yla,Maselli:2016gxk,Minamitsuji:2016hkk}. In computing the $\BI$--$\com$ relations we do not use equations of state containing quarks or hyperons, nor do we use particularly soft ones such as PAL2. 

\subsection{Neutron Stars}

We begin by examining the properties of Neutron stars.

\subsubsection{Maximum Mass}

Previously \cite{Babichev:2016jom}, we found that the maximum mass for neutron stars can be larger than $2M_\odot$ and may be larger than $3M_\odot$. The two equations of state we used in \cite{Babichev:2016jom} can hardly be considered generic and so in figure \ref{fig:MEOS} we plot the maximum mass for each equation of state. Each equation of state is described in Appendix \ref{app:EOS} and we remind the reader that they are a faithful representation of those used in the literature. Note that we do not include hyperonic or quark equations of state. Clearly, the trend of increasing maximum mass is ubiquitous and masses in the range $2M_\odot\le M\le3M_\odot$ are typical for $\U\lsim-0.03$. For GR, this is more like $1.5M_\odot\le M\le2M_\odot$. Interestingly, equations of state that are excluded in GR because they cannot account for the observed neutron star of $2M_\odot$ could be revived by modifying gravity. 

The maximum mass in GR is limited by a causality condition i.e. the condition that the sound speed should be $\dd P/\dd\rho\le1$ (see \cite{Koranda:1996jm,Lattimer:2010uk,Lattimer:2015nhk}), and so the observation of neutron stars with masses and radii that violate this may then point to alternative gravity theories since such objects cannot be accounted for in GR. In figure \ref{fig:MREOS} we plot the maximum mass and radius for each EOS, and, evidently, the majority indeed violate the GR causality condition when $\U\lsim-0.03$. It would be interesting to calculate the equivalent condition in beyond Horndeski theories but this lies well beyond the scope of the present work\footnote{The calculation in beyond Horndeski theories is not as straightforward as repeating the GR calculation using the mTOV equations. The scalar degree of freedom and matter are kinetically mixed, and so one must find the speed of scalar and density waves by diagonalising perturbations about the equilibrium structure (this is similar to what must be done to derive the speed of cosmological perturbations \cite{Gleyzes:2014qga}). Furthermore, the GR condition is derived by conjecturing that the equation of state $P=\rho-\rho_0$, for which $\dd P/\dd \rho=1$, produces maximally compact stars. It is not clear that this remains the case in beyond Horndeski theories since the mTOV equations contain new terms that depend on the derivative of the density. Such terms are absent in GR. Finding the maximally compact EOS would require a detailed numerical study similar to \cite{Rhoades:1974fn,Fujisawa:2015nda}.  }.

\begin{figure}
{\includegraphics[width=0.49\textwidth]{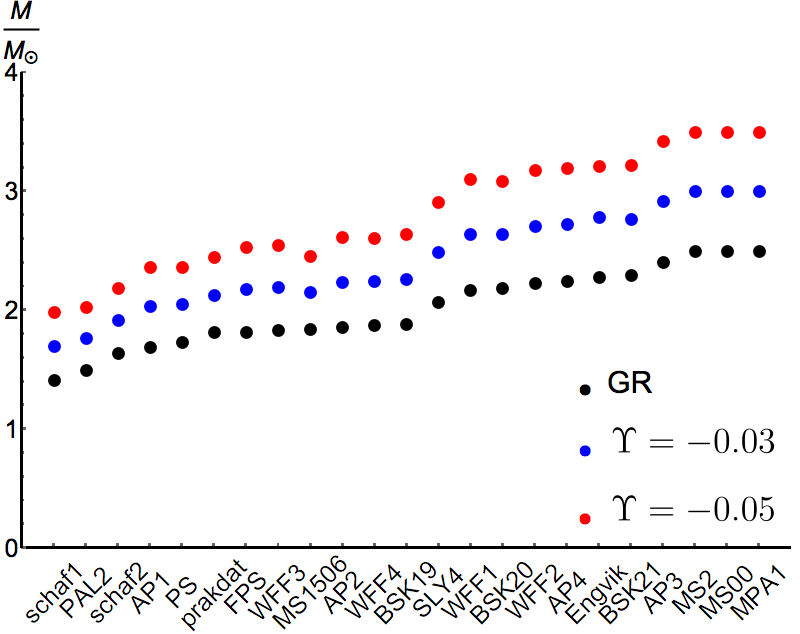}}
\caption{The maximum mass for each equation of state for values of $\U$ indicated in the figure.}\label{fig:MEOS}
\end{figure}
\begin{figure}
{\includegraphics[width=0.49\textwidth]{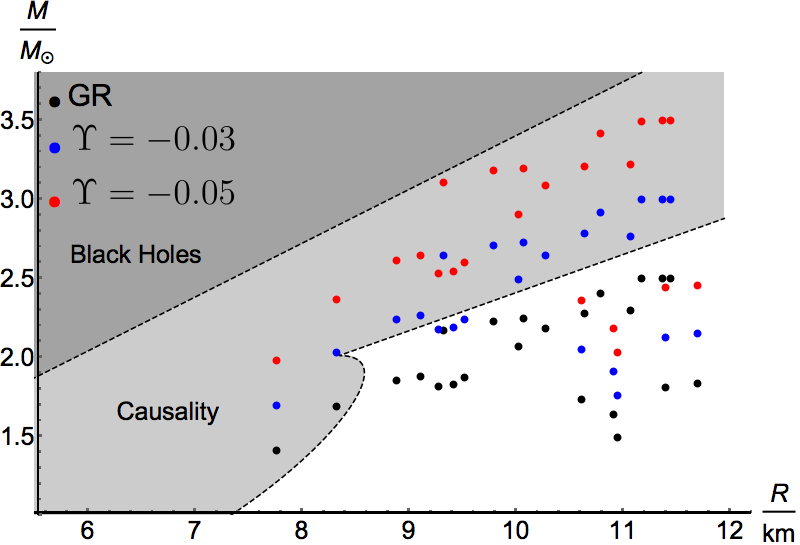}}
\caption{The maximum mass and radius for each equation of state. The values of $\U$ are the same as in figure \ref{fig:MEOS}. The light gray shaded region shows the condition for causality in GR i.e. the condition for the sound speed to be $\le1$ and assumes that the heaviest observed neutron star has a mass of $2.01M_\odot$. The dark gray region corresponds to objects that would be more compact than black holes i.e. $R<2\GN M$.}\label{fig:MREOS}
\end{figure}
\subsubsection{$\BI$--$\com$ Relations}\label{sec:IC}

\begin{figure*}[ht]
{\includegraphics[width=0.49\textwidth]{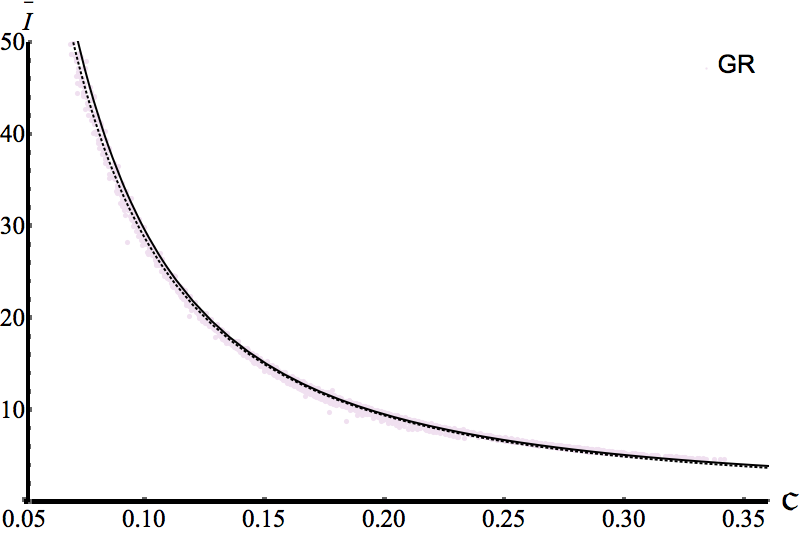}}
{\includegraphics[width=0.49\textwidth]{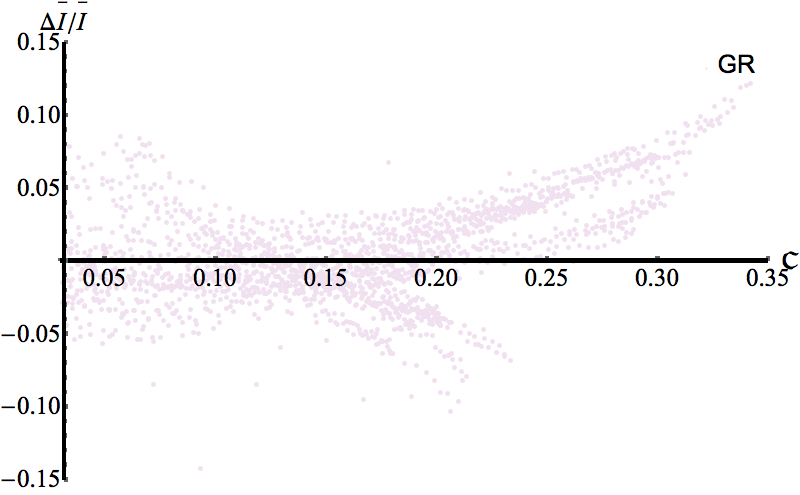}}
{\includegraphics[width=0.49\textwidth]{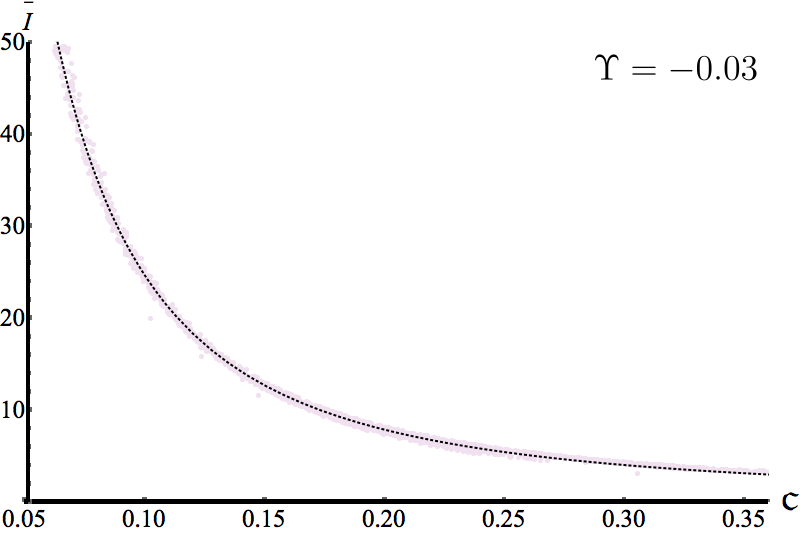}}
{\includegraphics[width=0.49\textwidth]{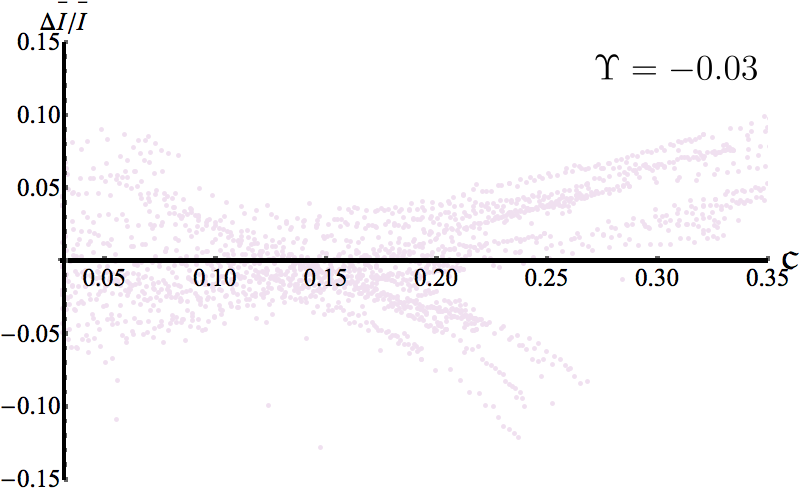}}
{\includegraphics[width=0.49\textwidth]{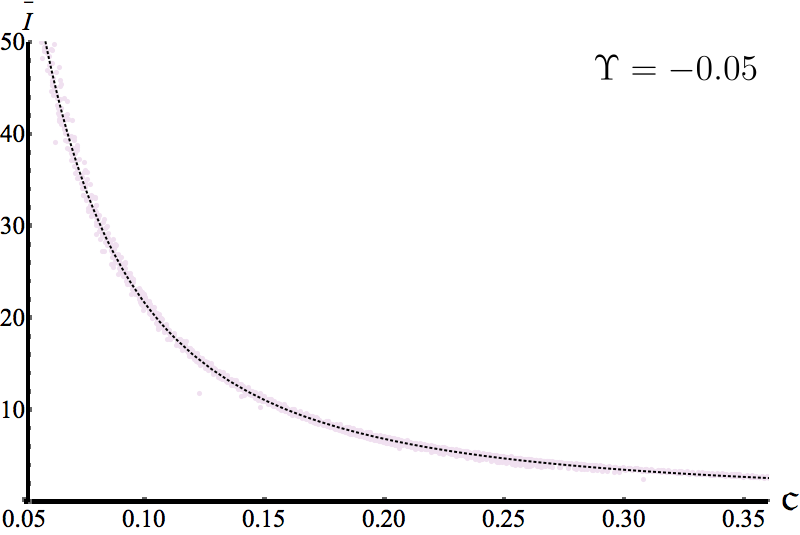}}
{\includegraphics[width=0.49\textwidth]{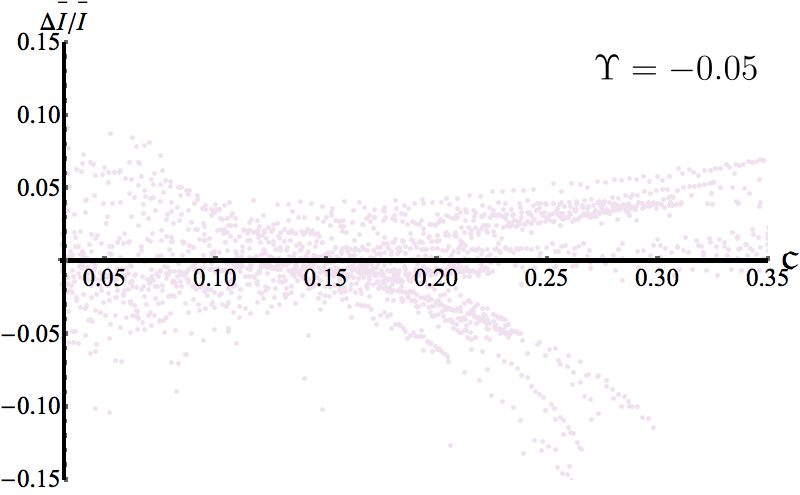}}
\caption{\emph{Left panels}: The $\BI$--$\com$ relations for GR (upper), $\U=-0.03$ (middle), and $\U=-0.05$ (lower). The Black solid line is the best fit of \cite{Breu:2016ufb} (upper panel only) and the black dashed line is our best fit. \emph{Right panels}: $\Delta \BI/\BI$ as a function of the compactness for GR (upper), $\U=-0.03$ (middle), and $\U=-0.05$ (lower). Each individual stellar model is represented by a purple dot in all cases.}\label{fig:GR}
\end{figure*}

As mentioned in the introduction, \cite{Breu:2016ufb} have found an approximately universal relation between the dimensionless moment of inertia $\bar{I}=Ic^2/\GN M^3$ and the compactness $\com=\GN M/R$ of the form
\begin{equation}\label{eq:fitform}
\BI= a_1\com^{-1}+a_2\com^{-2}+a_3\com^{-3}+a_4\com^{-4};
\end{equation}
In what follows, we will fit our modified gravity models to a relation of this form. In the upper left panel of figure \ref{fig:GR} we plot the $\BI$--$\com$ relation for individual stellar models for GR and show the best-fitting relation found by \cite{Breu:2016ufb} and our own, whose coefficients are given in table \ref{tab:as}. One can see that our relation agrees well with that of \cite{Breu:2016ufb}\footnote{We have not shown their best-fitting coefficients for clarity reasons but if one compares the two one finds small differences. This is to be expected since we use different equations of state and a different code to calculate the stellar models. What is important is that the two curves match very closely in the region $[0.05,0.40]$}. In the upper right panel we show the residuals $\Delta\BI/\BI=(\BI_{\rm fit}-\BI)/\BI$; one can see that these are less than 10\% and that there is no clear correlation with $\com$. We plot the equivalent figures for $\U=-0.03$ and $\U=-0.05$ in the middle and lower panels of figure \ref{fig:GR} and give the coefficients for the fitting functions in table \ref{tab:as}. Evidently, a similar (approximately) universal relation holds in both cases. 

\begin{figure}[ht]
{\includegraphics[width=0.49\textwidth]{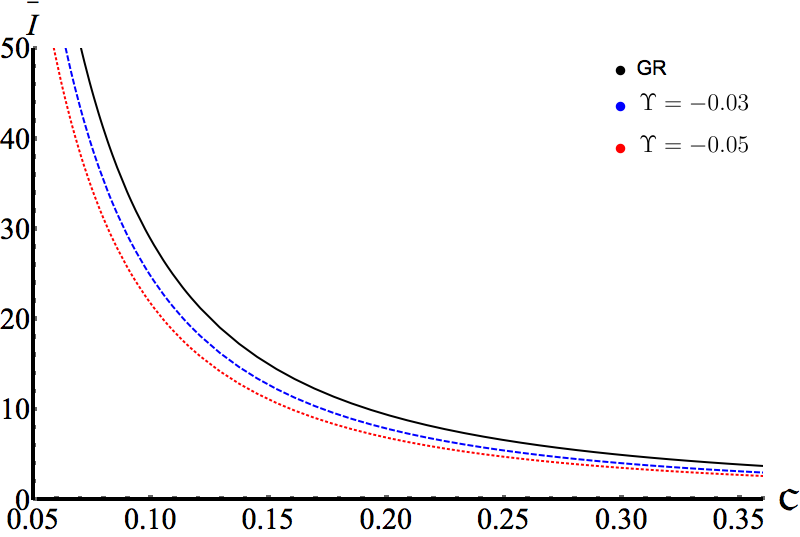}}
\caption{The $\BI$--$\com$ relation for GR (black solid curve) and beyond Horndeski theories with $\U=-0.03$ (blue, dashed) and $\U=-0.05$ (red, dotted).}
\label{fig:comb}
\end{figure}

\begin{figure}[ht]
{\includegraphics[width=0.49\textwidth]{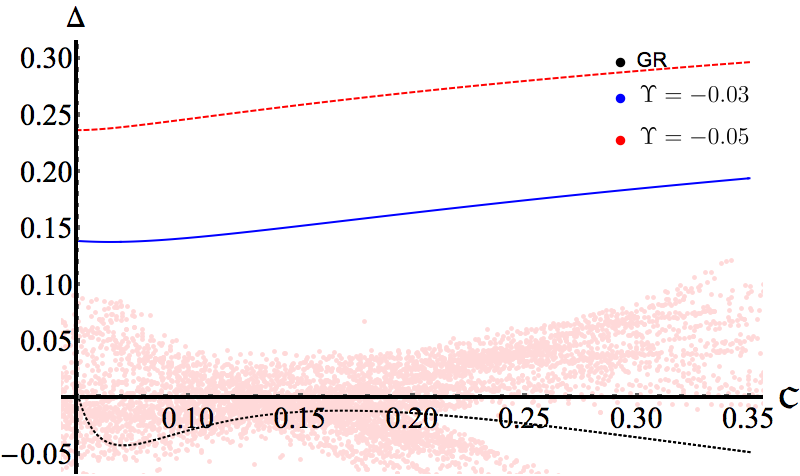}}
\caption{The fractional difference between the best-ftting $\BI$--$\com$ relations ($(\BI_{\U}(\com)-\BI_{\rm GR}(\com))/\BI_{\rm GR}(\com)$) for $\U=-0.03$ (blue, solid) and $\U=-0.05$ (red, dashed). We also show the fractional difference between our GR relation and the one found by \cite{Breu:2016ufb} (black, dotted) as well as the scatter in all three best-fitting relations (light red dots).}
\label{fig:comb2}
\end{figure}

The coefficients in the table by themselves are not particularly illuminating and a cursory glance does not reveal whether the differences between the relations for the different theories are significant or not. This is partly because the fitting function typically used is phenomenological and it is not clear how much degeneracy there is between the free parameters. For this reason, we have plotted two figures better suited to show that the differences between the GR and beyond Horndeski theories is significant. In figure \ref{fig:comb} we plot all three relations on the same axes. Evidently, there is a marked difference between the three. To quantify this, in figure \ref{fig:comb2} we plot the quantity $\Delta(\U,\com)\equiv (\BI_{\U}(\com)-\BI_{\rm GR}(\com))/\BI_{\rm GR}(\com)$, where $\BI_{\rm GR}$ is our best-fitting $\BI$--$\com$ relation for GR and $\BI_{\U}$ is the equivalent relation for a beyond Horndeski theory with parameter $\U$. We also plot the quantity $\Delta_{GR}\equiv(\BI_{\rm GR}-\tilde{I}_{\rm GR})/\tilde{I}_{\rm GR}$ where $\tilde{I}_{\rm GR}$ is the best-fit relation found by \cite{Breu:2016ufb}. We also plot the scatter in the best-fitting relation for all three theories. The difference between the GR relations is commensurate with the scatter in the best-fitting relations whereas the difference between the GR and beyond Horndeski relations is far greater than this ($\gsim 15\%$). Therefore a precise measurement of this relation has the power to discriminate between different theories. We note that many alternative theories of gravity, such as massless scalars coupled to matter and Einstein-Dilaton-Gauss-Bonnet, predict similar relations to GR \cite{Staykov:2016mbt,Maselli:2016gxk} and can therefore cannot be probed using the $\BI$--$\com$ relation.


\begin{table*}\centering
\begin{tabular}{c|c|c|c|c}
Theory & $a_1$ & $a_2$ & $a_3$ & $a_4$\\\hline
GR & $0.573221$ & $0.297835$ & $-7.56094\times10^{-3}$ & $7.92444\times10^{-5}$\\
$\U=-0.03$ & $0.366024$ & $0.277797$ & $-7.62076\times10^{-3}$ & $8.64847\times10^{-5}$\\
$\U=-0.05$ & $0.327218$ & $0.237928$& $-5.99316\times10^{-3}$&$5.89305\times10^{-5}$
\end{tabular}\caption{Coefficients for the fitting relation \eqref{eq:fitform}.}\label{tab:as}
\end{table*}

\subsection{Hyperon and Quark Stars}\label{sec:exotica}

Stars containing particles such as hyperons, kaons, or quarks in a colour-deconfined phase have been posited to exist, and their study is an active and ongoing area of research, and several of the equations of state we have used contain such particles. In this section we briefly discuss hyperonic and quark stars in beyond Horndeski theories, focusing on the hyperon puzzle and the transition from hyperon to quark stars. The former phenomenon can be solved by beyond Horndeski theories, whilst the latter remains a feature of theory, just as in GR. We note that our equations of state containing quarks are of the strange quark matter (SQM) form and are based on the MIT bag model \cite{Chodos:1974je}; they do not contain nucleons. Thus, when we refer to quark stars we refer to objects composed solely of SQM rather than neutron stars with quark cores.

\subsubsection{The Hyperon Puzzle and Quark Stars}

The hyperon problem (see \cite{SchaffnerBielich:2008kb,Vidana:2015rsa,Chatterjee:2015pua,Bombaci:2016xzl} for reviews) is a long-standing one in both the nuclear physics and neutron star communities. Nuclear theories predict that hyperons (baryons with non-zero strangeness), in particular the lowest mass $J=1/2$ baryon octet $\{n$, $p$, $\Sigma^{\pm}$, $\Sigma^0$, $\Xi^0$, $\Xi^-$, $\Lambda\}$, should appear at high densities. It is therefore inevitable that hyperons should contribute to the structure of neutron stars since densities far in excess of the hyperon threshold density can be reached in the core. Equations of state that include hyperons are substantially softer than pure nucleonic ones (see \cite{Bombaci:2016xzl} figure II), which has the consequence that the resulting maximum mass of neutron stars is significantly lower than stars containing nucleons only. In particular, the maximum mass predicted by realistic hyperonic equations of state \cite{Glendenning:1984jr,Glendenning:1991es,Schulze:2011zza} lies well below the heaviest presently observed neutron star mass of $2.01\pm0.04M_\odot$ \cite{Antoniadis:2013pzd}. This apparent tension between nuclear physics and neutron star astronomy constitutes the so-called \emph{hyperon puzzle}, and several potential resolutions within the realm of nuclear physics have been proposed.

The simplest explanation is our lack of understanding of hypernuclear physics. Hyperon interactions are poorly understood due in part to calculational difficulties and a lack of experimental data. Hyperon-nucleon and hyperon-hyperon interactions can be repulsive and can produce the additional pressure needed to support stars as heavy as $2M_\odot$ (see \cite{vanDalen:2014mqa} for example) although the values of the coupling constants in the theory are not presently known and are typically chosen in order to achieve the requisite $2M_\odot$. Similarly, one can scan the parameter space of effective theories including hyperons and find parameter choices that give $2M_\odot$ stars \cite{Oertel:2014qza}. In a similar vein, three body interactions (TBIs) are expected to be repulsive and may stiffen the equation of state (EOS) sufficiently to resolve the puzzle (see \cite{SchaffnerBielich:2008kb,Bombaci:2016xzl} and references therein). TBIs may also raise the density threshold for the appearance of hyperons so that they are simply not present in the cores of neutron stars \cite{Lonardoni:2014bwa}. Again, there is a lack of experimental data pertaining to TBIs and so whether they can resolve the puzzle remains to be seen. 

A more exotic solution is the presence of deconfined quark matter at high densities (see \cite{Glendenning:1997wn,Heiselberg:1999mq,Xu:2002wd} and references therein). In this scenario, there are two classes of compact stars: quark stars and hyperon stars, the latter being unstable above a threshold mass \cite{Berezhiani:2002ea,Berezhiani:2002ks,Drago:2004vu,Bombaci:2004mt,Fraga:2001id,Weissenborn:2011qu,Logoteta:2013aca,Orsaria:2013hna,Fraga:2013qra,Miyatsu:2015kwa}. Above this mass, a first-order phase transition to deconfined quark matter \cite{Hsu:1998eu} occurs in the core and the EOS becomes stiffer, resulting in stars of similar mass but smaller radii. The transition liberates a large amount of energy  ($10^{53}$ erg), which may be the source for gamma ray bursts. The reader is reminded that, in what follows, we model this transition using a pure SQM model only so that our quark star models contain no nucleons.  

\subsubsection{Hyperon and Quark Stars in Beyond Horndeski Theories}

\begin{figure*}[ht]
{\includegraphics[width=0.49\textwidth]{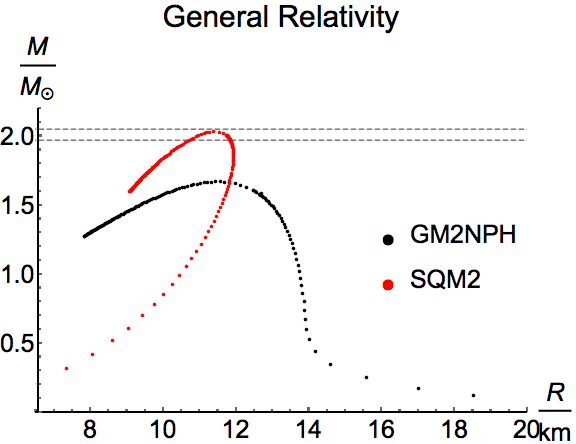}}
{\includegraphics[width=0.49\textwidth]{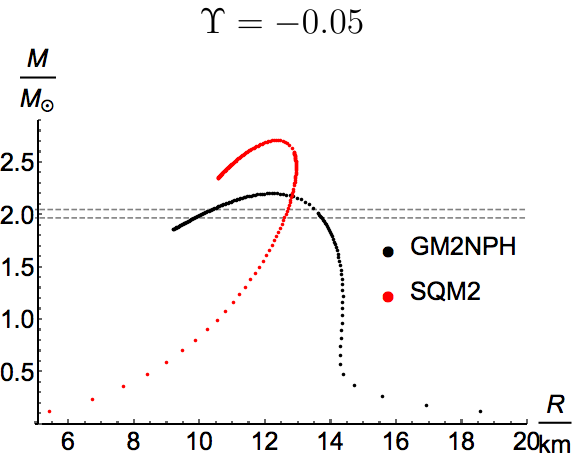}}
\caption{The mass-radius relation for hyperon (GM2NPH) stars (black) and quark (SQM2) stars (red) for general relativity (left panel) and beyond Horndeski theories with $\U=-0.05$ (right panel). The region between the gray dashed lines corresponds to the 1-$\sigma$ region for the heaviest mass object observed \cite{Antoniadis:2013pzd}. Note that the axes have different scales.}\label{fig:exotica}
\end{figure*}

\begin{figure}[h]
{\includegraphics[width=0.49\textwidth]{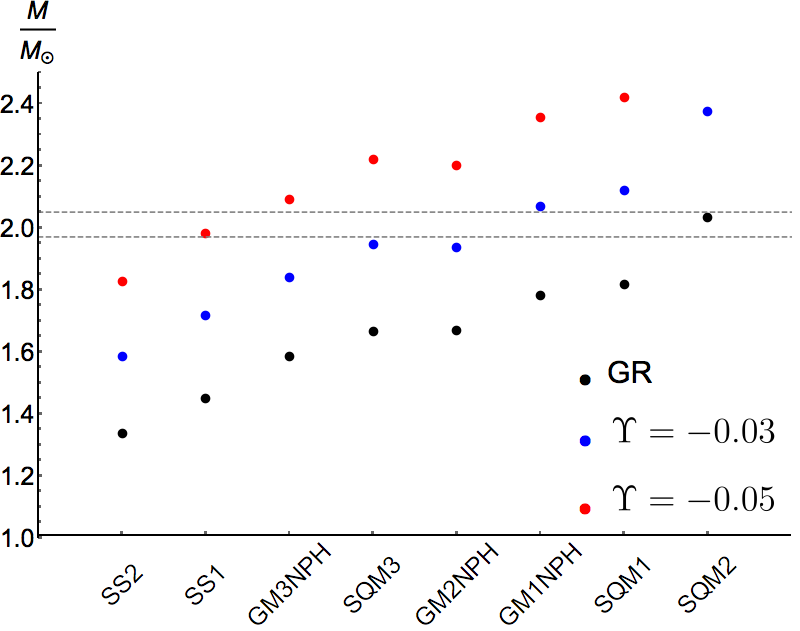}}
\caption{The maximum masses of hyperonic and quark stars in GR (black dots) and beyond Horndeski theories with $\U=-0.03$ (blue dots) and $\U=-0.05$ (red dots). The upper and lower bounds on the mass of the heaviest observed neutron star presently observed ($2.01\pm0.04M_\odot$ \cite{Antoniadis:2013pzd}) are shown using the grey dashed lines. }
\label{fig:QH}
\end{figure}

In figure \ref{fig:exotica} we plot the mass-radius relation for an equation of state containing hyperons, GN2NPH, and an equation of state containing quarks only, SQM2, for both general relativity and beyond Horndeski with $\U=-0.05$. Focusing on the left panel (GR), one sees that the maximum mass is around $1.5 M_\odot$ (one finds similar results for more modern equations of state \cite{Schulze:2011zza}), well below the mass of the heaviest compact object presently observed \cite{Antoniadis:2013pzd}. This is one manifestation of the hyperon problem. One can see that stable quark stars with masses compatible with this exist in the region where hyperon stars are unstable and so one resolution in GR is that stars with masses $\gsim 1.6M_\odot$ contain quark matter in a colour-deconfined state. The right panel shows the same curves but for beyond Horndeski theories with $\U=-0.05$. One can see that the maximum mass for hyperon stars is compatible with the observations of \cite{Antoniadis:2013pzd}, and therefore there is no hyperon puzzle in these theories \footnote{A similar result is found for a restricted class of $f(R)$ theories \cite{Astashenok:2014pua}.}. Interestingly, the situation with quark stars is similar in that more massive stable objects can exist in the region where hyperonic stars are unstable. This theory therefore predicts stable quark stars with masses in excess of $\sim2M_\odot$. Numerically, we find that the hyperon puzzle is resolved when $\U\lsim-0.04$, although the precise value depends slightly on the equation of state. We show the maximum masses for hyperon (GM equations of state) and quark stars (SS and SQM equations of state) in GR and beyond Horndeski theories in figure \ref{fig:QH}. One can see that the hyperon puzzle is more readily resolved in beyond Horndeski theories.

Given the large uncertainty in the hyperon equation of state, it is important to look for falsifiable predictions that can distinguish our resolution from other more conventional propositions. One promising test of these parameters is the $\bar{I}$--$\com$ discussed above since parameters that resolve the hyperon puzzle show clear deviations in their $\bar{I}$--$\com$ from the GR prediction. It would be interesting to examine this in more detail using more realistic equations of state but the lack of analytic fits for such models precludes this possibility for now\footnote{The appearence of $\dd P/\dd\rho$ and similar terms in the mTOV equations mandates the use of analytic fits rather than tabulated results. See \cite{Babichev:2016jom}.}.


\section{Discussion}\label{sec:disc}

In this section we discuss our findings above. We begin by summarising the main points before discussing the generalisation to more realistic models.

\subsection{Summary of Results}

Our main results can be summarised as follows:
\begin{itemize}
\item We have generically found that beyond Horndeski theories can give neutron star masses larger than $2M_\odot$ for small beyond Horndeski parameters $\Upsilon<0$ that are not constrained elsewhere; the maximum mass can be as large as $3M_\odot$ or even larger for more extreme parameter choices.
\item The $\bar{I}$--$\com$ relation persists in these theories and the difference from the GR relation is larger than the scatter due to different equations of state when $\U\lsim-0.03$. This relation therefore provides a robust equation of state-independent test of the theory provided that it persists in more realistic models. We will argue that this is likely to be the case below.
\item The hyperon puzzle is resolved in these theories because the maximum mass is enhanced by modifications of the TOV equation. Our proposed resolution can be falsified using other probes of the theory, for example, cosmological constraints on the EFT parameters $\alpha_i$ or tests of the $\BI$--$\com$ relation.
\item Quark stars are stable but should have masses in excess of $2M_\odot$.
\end{itemize}

\subsection{Measurements of the $\bar{I}$--$\com$ Relation}

Measurements of the $\bar{I}$--$\com$ relation are difficult because in order to find the moment of inertia one needs to measure the contribution of the (post-Newtonian) spin-orbit coupling to the precession of the periastron of the orbit of neutron star binaries, and so one requires a highly-relativistic system. A few such systems exist, and much attention has been paid to the double pulsar PSR J0737-3039A \cite{Burgay:2003jj,Lyne:2004cj}, for which the masses have already been measured to high precision using the properties of binary orbits. The spin-orbit coupling may be measured to $\sim10$\% accuracy \cite{Lattimer:2004nj} in the future and, accounting for the latest developments in the next generation of radio telescopes, it is possible that this measurement can be made in the next decade \cite{Kehl:2016mgp}. The Vainshtein mechanism is fully operative outside of the neutron stars and so the dynamics of this system are unchanged in beyond Horndeski theories. 

An alternative method of measuring the moment of intertia is to extract the tidal polarisability from the gravitational wave signal of neutron star-neutron star or neutron star-black hole mergers \cite{Yagi:2016bkt}. Such a measurement should be possible with advanced LIGO \cite{Newton:2016weo,Hotokezaka:2016bzh}, although the merger rate for such systems is currently unclear \cite{Abbott:2016ymx}. In this case, one may need to account for the varying speed of tensors predicted generically in beyond Horndeski theories \cite{Bettoni:2016mij}. 

Finally, one also needs to measure the neutron star radius in order to calculate the compactness. This can be achieved using X-ray observations \cite{Ozel:2016oaf} or by using universal $I$--$M$--$R$ relations \cite{Lattimer:2004nj}\footnote{In the latter case one would need to derive such relations for beyond Horndeski theories since the current relations are fit to GR solutions.}.

\subsection{Technical Considerations}

The {two} main constituents of our formalism were
\begin{enumerate}
\item The simple model presented in \eqref{eq:action}: a {k-essence} term and the beyond Horndeski quartic galileon.
\item Asymptotic de Sitter space-time.
\end{enumerate}
The combination of both constituents resulted in an exact Schwarzschild-de Sitter space-time outside the star and, furthermore, gave analytic relations between the coefficients $k_2,\,\zeta,\,f_4$, and $H$ that come from the Friedman equations because the scalar itself drives the asymptotic de Sitter solution. We now proceed to discuss each constituent in turn, focusing on how they may be relaxed in future studies.

The first was simply an aesthetic choice made for simplicity. {Note that the model (\ref{eq:action}) is a subclass of the more general ``three Graces'' family of the beyond Horndeski theory \cite{Babichev:2016kdt}, 
which allows for exact Schwarzschild-de-Sitter solution outside a star. Therefore, it is not difficult to perform similar calculations for the theories belonging to this three Graces family (in particular the work \cite{Cisterna:2015yla} studied another subclass of the three Graces family that includes the ``John''  term). On the other hand, one should be able to extend this study also to other beyond Horndeski action (outside of the three Graces) 
with little changing except the length of the resulting equations of motion.} The one technical complication that may arise is the following: The metric potentials and $\omega$ have boundary conditions at the centre and edge of the star whereas the scalar has boundary conditions at the centre and at infinity. In this simple model, it was possible to eliminate the scalar from the mTOV equations by solving an algebraic equation for $\dd\phi/\dd r$ that was found by integrating the equation of motion once. This is a result of the shift symmetry $\phi\rightarrow\phi + c$ which guarantees that the equation of motion is in the form of a current {conservation $\nabla_\mu j^\mu=0$}. More general theories that break the shift-symmetry will not have such a structure and so the scalar's equation of motion must be solved in conjunction with the other equations. Whilst not a debilitating obstacle, this will neccesitate the use of more complicated numerical boundary value techniques such as shooting or relaxation. One would also need to check numerically that the PPN parameters satsify $\gamma=\beta=1$ since it is likely that the Schwarzchild-de Sitter solution will be lost. Note that when this is the case one can still self-consistently determine the correct branch of solution using the technique outlined in \cite{Babichev:2016jom}, Appendix B.

The second constituent greatly simplified the computation because both $H$ and $\dot{\phi}(=v_0)$ were constant (see equation \eqref{eq:phids}). We do not live in an exact de Sitter Universe and so clearly future work should concentrate on deriving the equations for more general Friedmann--Lema\^\i tre--Robertson--Walker (FLRW) 
 backgrounds where $H$ and $\dot{\phi}$ depend on cosmic time. This is not an insurmountable challenge because it is not strictly necessary that $H$ and $\dot{\phi}$ are constant in order to derive the mTOV equations. Indeed, this has already been done in the weak-field limit for general FLRW backgrounds by \cite{Kobayashi:2014ida,Koyama:2015oma}. Importantly, there are some cosmological quantities that are not suppressed in the quasi-static sub-Horizon limit\footnote{By this, we mean that fields sourced by the star are time-independent. The existence of a sub-Horizon and quasi-static limit is necessary to have sensible physics in the solar system. Schwarzchild-de Sitter space-time is static and so only the sub-Horizon limit was important. The necessity of taking the quasi-static limit is a new feature of more general FLRW backgrounds.}. In our model, these were easy to identify because one has $f_4\sim\mpl^2/v_0^4$ (see \eqref{eq:f41}). Noting that $v_0^2\sim H^2\mpl^2/k_2\sim H^2\mpl^2$\footnote{Note that one can always absorb $k_2$ by performing a field redefinition and so we can take $k_2\sim\mathcal{O}(1)$ without loss of generality.}, we also have $f_4H^4\mpl^2\sim\mathcal{O}(1)$ and so quantities proportional to this survive in the sub-Horizon limit. Indeed, these quantities determine the value of $\U$. Moving to FLRW backgrounds, one can introduce the dimensionless parameter $g = f_4\dot{\phi}^4/\mpl^2$, which is $\mathcal{O}(1)$ and therefore survives in the sub-Horizon limit. In addition to this, one must make assumptions about $\ddot{\phi}$ and $\dot{H}$. Either they are negligible compared with $H\dot{\phi}$ and $H^2$ respectively, or they are of the same order. In the former case, one is left in a similar situation to that studied in this work and so similar equations are expected. In the latter case, there are new dimensionless parameters that survive in the sub-Horizon limit, for example, $f_4H^2\ddot{\phi}\mpl\sim g\sim\mathcal{O}(1)$. In this case, these can contribute to $\uo$ and $\ut$, as well as introduce new terms in the mTOV equations. Relaxing the restriction to a de Sitter background does not then render the derivation of the mTOV equations impossible, but it certainly makes the resulting formulae longer and more complicated. 

Another potential issue with more general FLRW space-times is potential strong bounds coming from the time-variation of $\GN$. When calculating the local value of $\GN$ measured in Cavendish-type experiments one finds that it is a combination of the constant $G=1/8\pi \mpl^2$ and the model parameters (see equation \eqref{eq:gdef}), in our case $H$. In the case of de Sitter space-time, $H$ is constant, but this is not the case for general FLRW space-times, and one generically expects $\GN$ to depend on $\dot{\phi}$ and similar quantities so that it is time-varying on the order of the Hubble time. Since galileon theories self-accelerate by utilising kinetic terms, the bound on the time-variation of $\GN$ ($\dot{\GN}/\GN<0.02H_0$ \cite{Williams:2004qba}) typically places tight restrictions on the model parameters \cite{Babichev:2011iz,Kimura:2011dc}, in contrast to quintessence-like models that accelerate using a scalar potential \cite{Sakstein:2014isa,Sakstein:2014aca,Ip:2015qsa,Sakstein:2015jca}. A detailed calculation of the resulting bound in beyond Horondeski theories is beyond the scope of this work but should be part of any analysis that attempts to move beyond de Sitter asymptotics.

\section{Conclusions}\label{sec:concs}

In this work we have studied slowly rotating relativistic compact objects in beyond Horndeski theories using 32 equations of state for neutron, hyperon, and quark matter. Our aim was to look for robust predictions that could be used to test said theories in the strong field regime. To this end, we considered a subset of beyond Horndeski theories corresponding to a k-essence term and a covariant quartic galileon. The equations of motion are incredibly long and complicated and so we restricted to the case where the space-time was asymptotically de-Sitter and driven by the scalar. Whilst this does not correspond to our present Universe, such simplifying assumptions are needed in order to make the problem tractable and, more importantly, one should verify that pursuing more realistic models is worthwhile. We have demonstrated here that it is. The equations of state we use have been found using several different calculational methods such as Hartree-Fock, mean field theory etc. and are therefore a representative sample of the equations of state found in the literature (see Appendix \ref{app:EOS} for a description of each equation of state.). 

We have derived the equations governing slow rotation, in particular, the coordinate angular velocity $\omega(r)$, and have solved them numerically to find the masses, radii, and moments of inertia of the resultant stars. We have shown that neutron stars with masses $>2M_\odot$ are predicted ubiquitously when $\U\lsim-0.03$ and that $3M_\odot$ can be obtained for stiffer equations of state that are currently favoured\footnote{Note that for extreme parameter choices, which are not currently ruled out, one can find even larger masses of order $5$ or $6M_\odot$. See \cite{Babichev:2016jom}.}. One can find stellar models that violate the causality limit in GR for certain equations of state. Such objects are robust tests of gravity beyond GR because they cannot be reproduced in GR while satisfying the GR causality limit, no mater the equation of state. It would be interesting to calculate the equivalent bound in beyond Horndeski theories---which changes because it results from solving the mTOV equations and an assumption about the maximally compact EOS---in order to verify that they are satisfied by our models, but we postpone such an analysis for future work due to technical complications and uncertainties about the maximally compact EOS in our theories.

Universal relations between the dimensionless moment of inertia $\bar{I}=Ic^2/\GN M^3$ and the compactness $\GN M/Rc^2$ that are independent of the equation of state have been found in GR \cite{Breu:2016ufb}, and other modified gravity theories \cite{Cisterna:2016vdx,Maselli:2016gxk}, which give similar relations. Here, we have shown that they are present in beyond Horndeski theories too. The beyond Horndeski relations are distinct from the GR prediction and therefore they can be used to test gravity in a manner that is independent of the equation of state and is hence free of the degeneracies associated with the current uncertainty in the nuclear equation of state. Measurements of the $\bar{I}$--$\com$ relation should be feasible within the next decade. 

We have also studied stars containing hyperons and quarks. Interestingly, beyond Horndeski theories can resolve the hyperon puzzle provided that $\U\lsim-0.04$, which is currently allowed by experimental constraints. More conventional resolutions focus on three body interactions (whose effects are largely unknown), or transitions to quark stars at high masses and so we have discussed whether a resolution in the form of beyond Horndeski theories is falsifiable. A precise measurement of the $\bar{I}$--$\com$ relation could rule out parameters where the hyperon problem is not present. Interestingly quark stars are still stable when the hyperon problem is resolved and both stable hyperon and quark stars can exist, the latter having masses $\gsim2M_\odot$.

We have ended by discussing our assumptions, paying attention to the potential to generalise our models to more realistic scenarios (general FLRW backgrounds) and whether or not we expect large qualitative changes in our results. We have argued that this is not the case. More general models may lack an exact analytic solution but one can still take an appropriate sub-Horizon and quasi-static limit making reasonable assumptions about the model parameters. In this case, the difference lies mainly in how $\U$ is related to more fundamental quantities. Our results clearly indicate that a study of more general backgrounds is warranted and such an investigation is left for future work. 

We have demonstrated here that a precise measurement of the $\bar{I}$--$\com$ relation could place tight constraints on this very general class of alternative gravity models. Indeed, such constraints would restrict possible cosmological deviations from GR and may act as consistency checks should non-zero values of $\alpha_H$ be preferred by upcoming cosmological surveys. 

\section*{Acknowledgements}

We are grateful to Emanuele Berti, Kenta Hotokezaka, Micaela Oertel, and Hector O. Silva for some very useful and enlightening discussions. Programme national de cosmologie et galaxies'' of the CNRS/INSU, France, and Russian Foundation for Basic Research Grant No. RFBR 15-02-05038. KK is supported by the UK Science and Technologies Facilities Council grants  ST/N000668/1 and the European Research Council through grant 646702 (CosTesGrav).

\bibliography{ref3}

\appendix

\section{Description of the Equations of State}\label{app:EOS}

In this appendix we briefly describe the equations of state we have used in this work. More detailed descriptions can be found in \cite{2008PhDT........15R}. The essential properties of each equation of state are given in table \ref{tab:EOSs}.

\begin{table}[h t]\centering
\begin{tabular}{c | c | c}
Name & Particle Content & Calculational Method\\\hline
BSK & $n$, $p$  &  Skyrme\\
APR, WFF, FPS & $n$, $p$ & Variational\\
PAL2, SLY4 & $n$, $p$ & Potential\\
MS, PRAKDAT & $n$, $p$ & Mean field theory\\
ENGVIK, PA1 & $n$, $p$ & Hartree-Fock\\ 
PS & $n$, $\pi^0$& Mean field theory \\
SCHAF & $n$, $p$, $K$ &  Mean field theory\\
GMNPH & $n$, $p$,  $K$, $H$ & Mean field theory\\
PCLNPHQ &  $n$, $p$, $H$, $Q$ &Mean field theory\\
SQM, SS & $Q$ & MIT bag model
\end{tabular}\caption{The equations of state used in our analyses. When one equation of state has several variants e.g. APR1, APR2 etc. we denote them collectively by type, in this example, APR. $K$ referes to Kaons, $H$ to the hyperons $\{n$, $p$, $\Sigma^{\pm}$, $\Sigma^0$, $\Xi^0$, $\Xi^-$, $\Lambda\}$, and $Q$ to the three lightest quarks $u$, $d$, and $s$. }\label{tab:EOSs}
\end{table}

\section{Derivation of the Main Equations}\label{sec:deriv}

In what follows, we will follow the derivation of \cite{Babichev:2016jom} whereby we treat the star as a static, spherically symmetric object embedded in de Sitter space. This allows us to include the cosmological time-dependence of the field on small scales and simultaneously have complete analytic control over the coordinate system without making any approximations or assumptions. This is because de Sitter space-time can be sliced in a Schwarzchild-like manner whereas more general cosmological backgrounds cannot. This property was vital to show that the Vainshtein breaking solution is indeed the physical one i.e. it has the correct asymptotics. The difference between the derivation here and the one in \cite{Babichev:2016jom} is that we use a different model ($c=0$).

\subsection{Cosmological Solution}

Since the theory is shift-symmetric ($\phi\rightarrow\phi+a$ for some constant $a$), the equation of motion for the scalar takes the form of a conservation equation for a conserved current $J^\mu$ i.e. $\nabla_\mu J^\mu=0$. Working in an  FLRW space-time
\begin{equation}
\label{metric_dS}
\dd s^2 = -\dd \tau^2 + a^2(\tau) \left(\dd\rpp^2 + \rpp^2\dd\Omega_2^2 \right) \,, 
\end{equation}
one finds that the only non-zero component is $J^0$ so that the equation of motion is $\partial_0(a^3J^0)$ implying that $J^0=0$\footnote{The solution with $J^0\propto a^{-3}$ decays rapidly.}. One then finds the scalar and Friedmann equations 
\begin{align}
J^0&=-k_2\pdot + (\z-12f_4)\pdot^3=0\quad\textrm{and}\\
3\mpl^2H^2&=k_2\pdot^2-3\left(\frac{\z}{2}-10f_4H^2\right)\pdot^4,\label{eq:fried}
\end{align}
which admit a de Sitter solution with $a(\tau)=e^{2H\tau}$ (with $H$ constant) and $\dot{\phi}=\pdot$ constant, for which
\begin{align}
k_2&=\frac{2\pdot^2}{3}\zeta-\frac{H^2}{4\pi G\pdot^2}\label{eq:k21}\\
f_4&=\frac{1}{48\pi G\pdot^4}+\frac{\zeta}{36 H^2}.\label{eq:f41}
\end{align}
These relations allow us to consistently relate the small-scale equations to the asymptotic cosmological solution. Note that we have defined $\mpl^2=1/8\pi G$ but $G$ is not the value of Newton's constant $\GN$ measured by Cavendish-style experiments. $G$ can be related to $\GN$ by deriving the weak-field limit and matching to the $00$-component of the PPN metric; we will do this below. 

\subsection{Stellar Structure Equations}

We now want to embed a static, spherically symmetric, and slowly rotating object into this space-time whose metric is given by
\begin{align}
\dd s^2& = -e^{\nu(r)}\dd t^2 + e^{\lambda(r)}\dd r^2 + r^2\dd \theta^2 +r^2\sin^2\theta\dd\phi^2 \nonumber\\&- 2\ve(\Omega-\omega(r))r^2\sin^2\theta\dd t \dd\phi\label{eq:HT2}
\end{align}
but we have the technical complication that \eqref{metric_dS} does not fit into this form. The coordinate transformation \cite{Babichev:2012re,Babichev:2016jom}
\begin{align}
\tau=t+\frac{1}{2H}\ln\left(1-H^2r^2\right)\quad\textrm{and}\quad\rpp=\frac{e^{-Ht}}{\sqrt{1-H^2r^2}}\, r\,
\end{align}
brings \eqref{metric_dS} into the form \eqref{eq:HT2} with $\ve=0$ and 
\begin{equation}
\nu(r)=-\lambda(r)=\ln\left(1-H^2r^2\right).
\end{equation}
The cosmological solution $\phi=\pdot\tau$ becomes
\begin{equation}
\phi = v_0t+\frac{v_0}{2H}\ln\left(1-H^2r^2\right).
\end{equation}
The object sources perturbations described by
\begin{align}
\label{asymp_dS2}
 \nu&= \ln\left(1-H^2r^2\right) + \delta\nu(r)+\oo(\ve^2),\\
  \lambda&= - \ln\left(1-H^2r^2\right)+\delta\lambda+\oo(\ve^2),\\
 \phi &= v_0t+\frac{v_0}{2H}\ln\left(1-H^2r^2\right) + \varphi(r)+\ve\varphi_1(r)+\oo(\ve^2)\quad\textrm{and}\label{eq:phi2}
\end{align}
where the metric potentials $\dn$ and $\dl$ do not receive corrections at $\oo(\ve)$ and $\varphi_1$ depends on $r$ only since it is a scalar under the rotation group. We therefore require that the quantities $\dn$, $\dl $, $\varphi$, and $\Omega-\omega$ vanish at large $r$. 

Next, we take the sub-horizon limit $Hr\ll1$ as we did previously for the model in \cite{Babichev:2016jom}. The new feature here is the new parameter $\zeta$, which from equation \eqref{eq:fried} scales as $\mpl^2H^2/\pdot^4$ and so we define an $\mathcal{O}(1)$ variable $s$ via
\begin{equation}
\zeta=\frac{6\mpl^2H^2}{\pdot^4}(1-s), 
\end{equation}
so that  equation \eqref{eq:k21} and \eqref{eq:f41} become 
\begin{align}
k_2&=\frac{H^2}{4\pi G \pdot^2}(1-2s)\label{eq:K22}\\
f_4&=\frac{1}{48\pi G \pdot^4}(2-s)\label{eq:f42}.
\end{align}
Note that since $k_2\propto H^2$ all $k$-essence vanish in the sub-Horizon limit. This is identical to our previous model ($c=1$) \cite{Babichev:2016jom} where $k_2\propto H^2$ and
\begin{equation}
f_4=\frac{1}{48\pi G\pdot^4}(1-\sigma^2);\quad\sigma= \frac{\Lambda}{3\mpl^2H_0^2}.
\end{equation}
In both cases only the terms proportional to $f_4$ contribute to the equations governing the structure of the object and the terms proportional to $k_2$ set the cosmology. We therefore have identical equations as our previous case with $\sigma^2\rightarrow s-1$. In particular, one has
\begin{align}
\GN&=\frac{3G}{5s-7}\label{eq:gdef}\\
\uo&=\ut=\U=-\frac{1}{3}\left(2-s\right)
\end{align}
and one requires $s>7/5$ in order for the Vainshtein breaking solution to exist, which implies that $\U>-1/5$. Importantly, the mTOV equations are identical to those given in appendix C in \cite{Babichev:2016jom} and so we do not repeat their derivation here.

The mTOV equations describe the structure of static objects and hence correspond to the parts of the equations of motion with $\ve=0$, or, alternatively, the $\oo(\ve)$ terms. The slow rotation is described by $\obar$, which appears at $\oo(\ve)$. The $\oo(\ve)$ part of the scalar  equation is solved by $\varphi_1=0$ and so the only change at this order is the new equation coming from the $t$--$\phi$ tensor equation, which is of the form 
\begin{equation}\label{eq:eeoffd}
G^t_\phi+H^t_\phi = 8\pi G T^t_\phi,
\end{equation}
where $H^\mu_\nu$ represents the contribution of the $f_4$ term to the tensor equation of motion. The energy-momentum tensor is
\begin{equation}
T^{\mu\nu} = (\rho+P)u^\mu u^\nu+P g^{\mu\nu},
\end{equation}
and
\begin{equation}
u^t = e^{-\frac{\dn}{2}}+\oo(\ve^2)\quad \textrm{and}\quad u^\phi=\ve\Omega e^{-\frac{\dn}{2}}+\oo(\ve^2).
\end{equation}
Substituting the zeroth-order solution for $\varphi$ \cite{Babichev:2016jom},
\begin{equation}
\frac{\varphi'^2}{\pdot^2}=\frac{re^{\dl-\dn}(\dl'+\dn')}{4(1+r\dn')},
\end{equation}
where the negative sign must be taken for the square root so that the space-time is asymptitcally de Sitter (\cite{Babichev:2016jom}), into equation \eqref{eq:eeoffd} and re-arranging, one finds the equation
\begin{equation}\label{eq:roteqn2}
\obar''= K_1(P,\rho,\dl,\dn,\U)\obar'+K_0(P,\rho,\dl,\dn,\U)\obar,
\end{equation}
where
\begin{equation}
K_1 = \frac{u_1}{v_1}\quad\textrm{and}\quad K_0=\frac{u_0}{v_0}
\end{equation}
with
\begin{widetext}
\begin{align}
u_1&=\dn'^4 r^4 \left(16 e^{2 \delta \nu }+5 \Upsilon _1\right)+\dl'^3 r^3 \Upsilon _1 (\dn' r+1)-\dn'^3 r^3 \left(80 e^{2 \delta \nu }+43 \Upsilon _1\right)\nonumber\\&+\dl'^2 r^2 \Upsilon _1 \left(7 \dn'^2 r^2+4 \dn'' r^2-9 \dn' r-20\right)-4 \dn'^2 r^2 \left(84 e^{2 \delta \nu }+\dl'' r^2 \Upsilon _1-19 \Upsilon _1\right)\nonumber\\&+4 \dn' r \left(-92 e^{2 \delta \nu }+3 \dl'' r^2 \Upsilon _1-5 \dn'' r^2 \Upsilon _1+24 \Upsilon _1\right)-16 \left(-\dl'' r^2 \Upsilon _1-5 \dn'' r^2 \Upsilon _1+8 \left(e^{2 \delta \nu }+\Upsilon _1\right)\right)\nonumber\\&+\dl' r \left(\dn'^3 r^3 \left(16 e^{2 \delta \nu }+11 \Upsilon _1\right)+4 \dn' r \left(12 e^{2 \delta \nu }+\dl'' r^2 (-\Upsilon _1)+\dn'' r^2 \Upsilon _1-2 \Upsilon _1\right)\right.\nonumber\\&\left.+4 \left(4 e^{2 \delta \nu }+\dl'' r^2 (-\Upsilon _1)-9 \dn'' r^2 \Upsilon _1+24 \Upsilon _1\right)+\dn'^2 r^2 \left(48 e^{2 \delta \nu }-53 \Upsilon _1\right)\right),\\
v_1&=2 r (\dn' r+1) \left(\dn''^2 r^2 \Upsilon _1+\dn'^2 r^2 \left(16 e^{2 \delta \nu }+\Upsilon _1\right)+2 \dn'' r \Upsilon _1 (\dn' r-4)\right.\nonumber\\&\left.+8 \dn' r \left(4 e^{2 \delta \nu }-\Upsilon _1\right)+16 \left(e^{2 \delta \nu }+\Upsilon _1\right)\right),\\
u_0&=256 \pi  G_N (5 \Upsilon _1+1) e^{\delta \lambda +2 \delta \nu } (P+\rho ) (r\dn' +1)^2\quad\textrm{and}\\
v_0&=\dl'^2 r^2 \Upsilon _1+\dn'^2 r^2 \left(16 e^{2 \delta \nu }+\Upsilon _1\right)+2 \dl' r \Upsilon _1 (\dn' r-4)+8 \dn' r \left(4 e^{2 \delta \nu }-\Upsilon _1\right)\nonumber\\&+16 \left(e^{2 \delta \nu }+\Upsilon _1\right).
\end{align}
\end{widetext}
We note that outside the star where $P=\rho=0$ and the space-time is Schwarzchild-de Sitter we have
\begin{equation}
\obar''+\frac{4}{r}\obar'=0,
\end{equation}
which is the same equation one finds in GR \cite{Hartle:1967he}.

\end{document}